\documentclass[aps,pra,twocolumn,amsmath,groupedaddress]{revtex4-2}
\usepackage{graphicx}
\usepackage{dcolumn}
\usepackage{bm}
\usepackage{color} 
\usepackage[tight]{subfigure}
\usepackage{hyperref}

\newcommand{\ket}[1]{\left|{}#1 \right>}
\newcommand{\bra}[1]{\left<{}#1 \right|}
\newcommand{\expect}[1]{\left<{}#1\right>}
\newcommand{\interproduct}[2]{\langle {}#1 | {}#2 \rangle}

\newcommand{\tr}[2]{\mathrm{tr}_{{}#2} \left({}#1\right)}

\begin{document}
	
	\title{Quantum walk on the Bloch sphere}
	\author{Liwei Duan}
	\email{duanlw@gmail.com}
	\affiliation{Department of Physics, Zhejiang Normal University, Jinhua 321004, China}
	
	\date{\today}
	
	\begin{abstract}
		A scheme for implementing the discrete-time quantum walk on the Bloch sphere is proposed, which is closely related to the SU(2) group. A spin cluster serves as the walker, whereas its location on the Bloch sphere is described by the spin coherent state. An additional spin that interacts with the spin cluster plays the role of a coin, whose state determines the rotation of the spin cluster. The Wigner function is calculated to visualize the movement of the walker on the Bloch sphere, with which the probability distribution and the standard deviation are also achieved. The quadratic enhancement	of variance for the quantum walk on the Bloch sphere is confirmed. Compared to the ideal quantum walk on a circle, the walker's states on the Bloch sphere are generally nonorthogonal, whose drawbacks can be eliminated by increasing the number of spins in the spin cluster.
	\end{abstract}

	\maketitle
	
	\section{Introduction}
	
	As a quantum counterpart to the classical random walk, the quantum walk has been widely employed in numerous realms, ranging from physics to computer science \cite{aharonov_quantum_1993,doi:10.1080/00107151031000110776,venegas-andraca_quantum_2012,manouchehri_physical_2014}. One of the most surprising features of the quantum walk is a quadratic enhancement of variances and possible exponential algorithmic speedups due to the quantum interference \cite{doi:10.1080/00107151031000110776}.  On one hand, the quantum walk provides a versatile platform to simulate physical phenomena, such as the nontrivial topological phase \cite{kitagawa_exploring_2010,kitagawa_observation_2012,ramasesh_direct_2017,xie_topological_2020}, non-Hermitian system \cite{mochizuki_explicit_2016,xiao_observation_2017,wang_observation_2019,xiao_non-hermitian_2020}, Anderson localization \cite{ahlbrecht_disordered_2011}, strongly correlated quantum matter \cite{preiss_strongly_2015}, dynamic quantum phase transitions \cite{wang_simulating_2019}, quantum-to-classical transition \cite{travaglione_implementing_2002,PhysRevA.67.042305,brun_quantum_2003,karski_quantum_2009}, etc.
	One the other hand, the quantum walk plays a significant role in quantum information, as it provides a powerful technique for building quantum algorithms and serves as a universal platform for quantum computation \cite{portugal_quantum_2013}.
	
	The implementation of quantum walks has been proposed or realized in different physical systems \cite{manouchehri_physical_2014}, such as the ion trap\cite{travaglione_implementing_2002,xue_quantum_2009,schmitz_quantum_2009,zahringer_realization_2010,Matjeschk_2012}, NMR \cite{ryan_experimental_2005}, CQED \cite{PhysRevA.67.042305}, nitrogen-vacancy centers in diamond \cite{hardal_discrete-time_2013}, the optical lattice\cite{dur_quantum_2002,preiss_strongly_2015}, single photon \cite{broome_discrete_2010,xiao_observation_2017}, single optically trapped atoms \cite{karski_quantum_2009}, Bose-Einstein Condensate \cite{dadras_quantum_2018,xie_topological_2020}, etc.
	Theoretically, they can be broadly classified into two categories: the discrete-time quantum walk \cite{aharonov_quantum_1993}, in which the walker propagates on a lattice in discrete time steps determined by an additional coin, and the continuous-time quantum walk \cite{farhi_quantum_1998}, in which the dynamics is totally governed by a time-independent lattice Hamiltonian. 
	This paper is mainly concerned with the former case, which is first introduced by Aharonov \textit{et al}. \cite{aharonov_quantum_1993}. 
	
	Generally, the discrete-time quantum walk consists of a walker moving in some space, and a flipped coin whose state determines the movement of the walker. The position space \cite{travaglione_implementing_2002,karski_quantum_2009,Matjeschk_2012}, momentum space \cite{dadras_quantum_2018,xie_topological_2020} and phase space \cite{PhysRevA.67.042305,schmitz_quantum_2009,Omanakuttan_2018} have been chosen as a platform for the walker to move. 
	In the phase space, one usually employs a harmonic oscillator as the walker, which is closely related to the Heisenberg-Weyl group \cite{PhysRevA.67.042305}. 
	The walker's state determines the location on the phase plane, which consists of all possible values of position and momentum variables. Previous theoretical studies mainly focus on ideal localized states for the walker in the phase space. There is no overlap, namely that states corresponding to different locations are orthogonal. However, one can hardly generate the orthogonal localized states for the walker physically, whereas nonorthogonal Gaussian states, such as the bosonic coherent state, are more feasible in the experiments \cite{PhysRevA.67.042305,schmitz_quantum_2009,matjeschk_quantum_2012,Matjeschk_2012,wineland_experimental_1998}.  The influences of  nonorthogonal walker's states have been studied, which can smear out the probability distributions \cite{Matjeschk_2012} and model transport processes in complex systems \cite{matjeschk_quantum_2012}.
	
	In this paper, I consider the quantum walk on the Bloch sphere. The Bloch sphere is a geometrical representation for systems closely related to the SU(2) group \cite{RevModPhys.62.867,doi:https://doi.org/10.1002/9783527628285.ch6}. The walker can be a cluster of spins, an angular momentum, or a coupled two-mode field through the Schwinger realization, whereas its location on the sphere can be described by the spin coherent state \cite{arecchi_atomic_1972,RevModPhys.62.867,gerry_remarks_2001}. The spin coherent state, also known as atomic or Bloch coherent state \cite{doi:https://doi.org/10.1002/9783527628285.ch6}, was introduced in the early 1970s by Radcliffe \cite{radcliffe_properties_1971}, Gilmore \cite{gilmore_geometry_1972,arecchi_atomic_1972} and Perelomov \cite{perelomov_coherent_1972}.  It has been widely employed to study the cooperative phenomena \cite{doi:https://doi.org/10.1002/9783527628285.ch6}, such as the superradiant phase transition, quantum magnetism and so on. In addition, the spin coherent state is an essential ingredient to construct the spin cat state and spin compass state \cite{davis_wigner_2021,akhtar_sub-planck_2021}. 
	Like the bosonic coherent state, the spin coherent states are generally nonorthogonal and can be generated in the experiments \cite{arecchi_atomic_1972,RevModPhys.62.867}. A visual description to the walker's states on the Bloch sphere can be achieved by calculating the Wigner function \cite{stratonovich_distributions_1957,varilly_moyal_1989,davis_wigner_2021,akhtar_sub-planck_2021}. 
	
	The paper is structured as follows. In Sec. \ref{sec:QW}, I revisit the basic properties of the spin coherent state and the Bloch sphere. Then, a physical implementation of quantum walk on the Bloch sphere is proposed. In Sec. \ref{sec:result}, I calculate the probability distribution and the standard deviation based on the Wigner function. An ideal quantum walk on a circle with orthogonal walker's state is also present for comparison. A brief summary is given in Sec. \ref{sec:conclusion}.

	\section{Quantum Walk on the Bloch sphere} \label{sec:QW}
	
\begin{figure}[htb]
		\centering
		\includegraphics[scale=1]{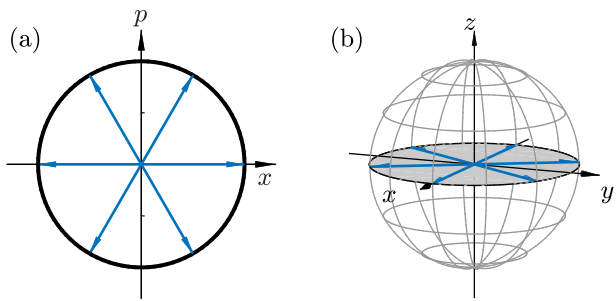} 
		\caption{Sketch of quantum walk on a circle with $L=6$ sites. (a) A harmonic oscillator on the phase plane; (b) a spin cluster  on the equator of the Bloch sphere. The blue arrows point to the possible locations of the walker.}\label{fig:phase}
	\end{figure}

	Previous studies on the quantum walk over a circle in the phase space mainly focus on the harmonic oscillator \cite{PhysRevA.67.042305,schmitz_quantum_2009}. The phase space corresponds to a plane consisting of all possible values of position and momentum variables as shown in Fig. \ref{fig:phase} (a), which associates with the Heisenberg-Weyl group. In this paper,  the quantum walk in phase space is extended to the Bloch sphere based on a spin cluster as shown in Fig. \ref{fig:phase} (b), which corresponds to the SU(2) group. Specifically, the spin cluster consists of identical spins which are permutational invariant.
	
	\subsection{Spin-coherent state and Bloch sphere}
	
	I begin by briefly reviewing the spin coherent state, which corresponds to a point on the surface of the Bloch sphere \cite{RevModPhys.62.867,doi:https://doi.org/10.1002/9783527628285.ch6}.
	It can be written as
	\begin{eqnarray}
		\ket{\theta, \phi} &=& \exp \left[\frac{\theta}{2} \left(e^{i \phi} \hat{J}_- - e^{-i \phi} \hat{J}_+\right)\right] \ket{J,J} \nonumber\\
		&=& \cos^{2J} \left(\frac{\theta}{2}\right) \exp \left[\tan \left(\frac{\theta}{2}\right) e^{i \phi} \hat{J}_-\right] \ket{J, J} , \label{eq:spin-coherent}
	\end{eqnarray}
	where $\hat{J}_{\gamma}$ ($\gamma=x,y,z$)  are the collective spin operators and $\hat{J}_{\pm} = \hat{J}_x \pm i \hat{J}_y$ are the corresponding ladder operators. $\hat{J}_{\gamma}$ can also be regarded as the generators of the SU(2) group. $\ket{J, J}$ is a Dicke state which satisfies $\hat{J}_z \ket{J, J} = J \ket{J, J}$.
	For the  spin coherent state $\ket{\theta, \phi}$, the expectation values of the collective spin operators $J_{\gamma} = \bra{\theta, \phi} \hat{J}_{\gamma} \ket{\theta, \phi}$ are
	\begin{eqnarray}
		\left(J_x, J_y, J_z\right)/J = \left(\sin \theta \cos \phi, \sin \theta \sin \phi, \cos \theta \right) . 
	\end{eqnarray}
	Therefore, it is located on the Bloch sphere with polar angle $\theta$ and azimuthal angle $\phi$ \cite{RevModPhys.62.867}. 
	It should be noted that \cite{arecchi_atomic_1972}
	\begin{eqnarray}
		\left| \interproduct{\theta, \phi}{\theta', \phi'}\right| = \cos^{2 J} \frac{\Theta}{2} , \label{eq:over_lap1}
	\end{eqnarray}
	where $\Theta$ is the angle between the $(\theta, \phi)$ and the $(\theta', \phi')$ directions and satisfies
	\begin{eqnarray}
		\cos \Theta = \cos \theta \cos \theta' + \sin \theta \sin \theta' \cos \left(\phi - \phi'\right) . \label{eq:over_lap2}
	\end{eqnarray} 
	Then, the spin coherent states are, in general, not orthogonal except for antipodal points ($\Theta=\pi$) \cite{RevModPhys.62.867}. The orthogonality is achieved in the limit of $J \rightarrow \infty$, for arbitrary two spin coherent states with $\Theta \ne 0$. 
	
	A generic rotation on the Bloch sphere can be described by the rotating operator, defined by
	\begin{eqnarray}
		\hat{R}_{\mathbf{n}}(\alpha) = e^{-i \alpha \mathbf{n} \cdot \hat{\mathbf{J}}}, 
	\end{eqnarray}
	which indicates a rotation by angle $\alpha$ along the $\mathbf{n}$ direction.
	Without loss of generality, $\mathbf{n}=(0,0,1)$ is chosen and the corresponding rotating operator is labeled as $\hat{R}_z (\alpha)$ in what follows.
	For each $\theta$, there exists a corresponding circle on the Bloch sphere. A set of equally displacing sites on the circle can be written as $\ket{\theta , \phi_n = n \delta \phi}$,
	with $n \in \left[- \frac{L}{2}, \frac{L}{2}\right]$, $\delta \phi = 2 \pi / L$, and $L$ as the total number of sites. From Eqs. (\ref{eq:over_lap1}) and (\ref{eq:over_lap2}), the overlap between different states $|\interproduct{\theta,\phi_m}{\theta, \phi_n}|$ is smallest and the quantum walk on the Bloch sphere can better mimic the ideal one when $\theta = \pi / 2$. Therefore, I focus on $\theta=\pi/2$ and the corresponding states are labeled as
	\begin{eqnarray}
		\ket{\phi_n} = \ket{\theta=\frac{\pi}{2}, \phi=n \delta \phi} ,
	\end{eqnarray} 
	which satisfy
	\begin{eqnarray}
		\left| \interproduct{ \phi_m}{ \phi_n}\right| = \left[ \frac{\cos (m - n) \delta \phi + 1}{2}\right]^{J} . \label{eq:over_lap3}
	\end{eqnarray}
	As an example, Fig. \ref{fig:phase} (b) depicts  $6$ sites on the equator, which correspond to $\theta = \pi / 2$ and $ \delta \phi = \pi / 3$. In the next section, I will proposed a scheme for implementing the quantum walk on such a kind of circular trajectory.

	\subsection{Physical implementation of the quantum walk on the Bloch sphere}
	
	Now I consider a universal model composed of two subsystems, which are described by the collective spin operators $\hat{\mathbf{J}}$ and $\hat{\mathbf{S}}$. One subsystem ($\hat{\mathbf{J}}$) serves as a walker, whereas the other one ($\hat{\mathbf{S}}$) serves as a coin whose state determines the movement of the walker.
	The total Hamiltonian can be written as
	\begin{eqnarray}
		\hat{H}(t) &=& \hat{H}_0 + \hat{H}_1 (t), \\
		\hat{H}_0 &=& 2 \kappa \hat{J}_z \otimes \hat{S}_z, \\
		\hat{H}_1 (t) &=& \sum_{k=0}^{+\infty} \hat{I}_{\mathrm{w}} \otimes \mathbf{h} \cdot \hat{\mathbf{S}} \delta  \left(t - k T\right) ,
	\end{eqnarray}
	where $\kappa$ is the interacting strength between two subsystems, $\mathbf{h} = \left(h_x, h_y, h_z\right)$ corresponds to a pulse acted on the coin with period $T$ and amplitude $h = \sqrt{h_x^2 + h_y^2 + h_z^2}$ along the direction $\mathbf{h} / h$,  and $\hat{I}_{\mathrm{w}}$ is the identity matrix of the walker. Such a kind of Hamiltonian commonly appears in various systems, such as atom-light interaction systems \cite{kuzmich_atomic_1998,bao_spin_2020,huang_dynamic_2021}, Bose–Einstein condensates \cite{jing_split_2019,huang_dynamic_2021} and magnetic clusters \cite{PhysRevA.71.042303}, etc. In this paper, I take two subsystems as spin clusters, whereas other systems can be dealt with accordingly. In terms of the Pauli matrices $\hat{\sigma}_{\gamma}$ , the collective spin operators can be written as $\hat{J}_{\gamma} = \sum_{i=1}^{N} \hat{\sigma}_{i, \gamma} / 2$ and $\hat{S}_{\gamma} = \hat{\sigma}_{\gamma} / 2$, where $N$ is the total number of spins in the spin cluster of the walker.
	
	
	The time evolution over one period is determined by 
	\begin{eqnarray}
		\hat{U}(T) = \hat{M} \cdot \hat{C},
	\end{eqnarray}
	with 
	\begin{eqnarray}
		\hat{M} &=& \exp \left(-i \hat{H}_0 T\right)  \label{eq:R_S}\\
		&=& \hat{R}_z (\kappa T) \otimes \ket{\uparrow} \bra{\uparrow} + \hat{R}_z (- \kappa T) \otimes \ket{\downarrow} \bra{\downarrow} , \nonumber\\
		\hat{C} &=& \exp \left(-i \hat{I}_{\mathrm{w}} \otimes \mathbf{h} \cdot \hat{\mathbf{S}}\right) \\
		&=& \hat{I}_{\mathrm{w}} \otimes \exp\left(-i \mathbf{h} \cdot \hat{\mathbf{S}}\right) . \nonumber
	\end{eqnarray}
	In each step of the quantum walk, one flips the coin and changes its state at first, which is determined by the coin-flip operator. Then, the walker shifts its location according to the coin's state, which is determined by the conditional-shift operator. Based on the time evolution operator $\hat{U}(T)$, one can find that $\hat{C}$ leads to a rotation of the coin state by angle $h$ along the $\mathbf{h} / h$ direction, which plays a role of the coin-flip operator.  One of the most frequently employed coin-flip operators is the Hadamard gate $\hat{H}_{\mathrm{c}}$, with
	\begin{eqnarray}
		\hat{H}_{\mathrm{c}} = \frac{1}{\sqrt{2}} \left(
		\begin{array}{cc}
			1 & 1 \\
			1 & -1
		\end{array}
		\right) .
	\end{eqnarray}
	The Hadamard gate up to a global phase factor is achieved
	\begin{eqnarray}
		\hat{C} = -i \hat{I}_{\mathrm{w}} \otimes \hat{H}_{\mathrm{c}} ,
	\end{eqnarray}
	by setting $\mathbf{h} = \left(\pi, 0, \pi\right) /\sqrt{2}$.
	$\hat{M}$ can be regarded as a conditional-shift operator. The interacting strength $\kappa$ and period $T$ are chosen such that $\kappa T = \delta \phi = 2 \pi / L$, which  leads to the transfer of  walker's states depending on the coin, namely
	\begin{eqnarray}
		\hat{M} \ket{\phi_n} \otimes \ket{\uparrow} = \ket{\phi_{n + 1}} \otimes \ket{\uparrow} , \nonumber\\
		\hat{M} \ket{\phi_n} \otimes \ket{\downarrow} = \ket{\phi_{n - 1}} \otimes \ket{\downarrow} .
	\end{eqnarray}
	
	
	Given that the initial state is $\ket{\psi(0)} = \ket{w} \otimes \ket{c}$,  with the walker and the coin initially at $\ket{w}$ and $\ket{c}$ respectively, the final state after $k$ steps would be 
	\begin{eqnarray}
		\ket{\psi(k)} = \left(\hat{M} \cdot \hat{C} \right)^k \ket{\psi(0)}.
	\end{eqnarray}
	In what follows, the initial state is set to be $\ket{w} = \ket{ \phi_0}$   and $\ket{c} = \ket{\uparrow}$. If the overlap (Eq. (\ref{eq:over_lap3})) between different walker's states is ignored, the quantum walk on the Bloch sphere reduces to an ideal one with orthogonal walker's states \cite{travaglione_implementing_2002}.

	\section{Results and Discussions} \label{sec:result}

	\begin{figure*}[htb]
		\centering
		\includegraphics[scale=0.95]{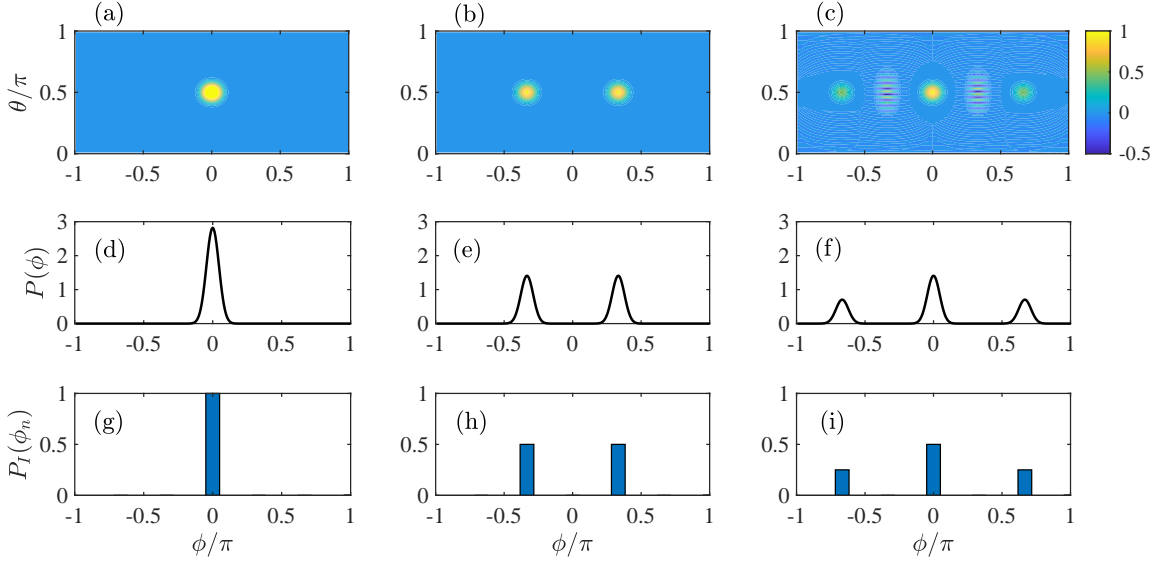} 
		\caption{Quantum walk on a circle with $L=6$ for the first two steps. (a), (d), and (g) correspond to the initial states ($k=0$). (b), (e), and (h) correspond to the states after one ($k=1$) step. (c), (f), and (i) correspond to the states after two ($k=2$) steps. (a), (b), and (c) refer to the Wigner functions.  (d), (e), and (f) refer to the probability distribution of the quantum walk for a spin cluster with $N=50$, whereas (g), (h), and (i) show the corresponding probability distribution of an ideal quantum walk with orthogonal walker's states. }\label{fig:PD}
	\end{figure*}
	
	The quantum walk is known for its ballistic spread quadratically faster than its classical counterpart which shows a diffusive spread. 	
	Because of the quantum interference effect, the variance of the quantum walk grows quadratically with the number of steps $k$ ($\sigma^2 \propto k^2$), compared to the linear growth ($\sigma^2 \propto k$) for the classical random walk.
	
	 In order to demonstrate the quadratic enhancement, one needs first calculate the probability distribution. For an ideal quantum walk, different walker's states are orthogonal. One can easily achieve the probability distribution as follows:
	\begin{eqnarray}
		P_I (\phi_n) = \bra{\phi_n} \hat{\rho}_{\mathrm{w}} (k) \ket{ \phi_n}, 
	\end{eqnarray}
	where $\hat{\rho}_{\mathrm{w}} (k) = \tr{\ket{\psi (k)} \bra{\psi (k)}}{c}$ is the reduced density matrix of the walker. Then, the standard deviation is given by
	\begin{eqnarray}
		\sigma_I = \sqrt{\expect{\phi^2} - \expect{\phi}^2} , \label{eq:var}
	\end{eqnarray}
	with $\expect{\phi^l} = \sum_n P_I (\phi_n) \phi_n^l$. 
	
	However, the quantum walk on the Bloch sphere corresponds to a set of spin coherent states, which are generally nonorthogonal, as indicated in Eq. (\ref{eq:over_lap3}). Fortunately, the Wigner function can be viewed as a quantum analogy to the classical probability density, which is able to visualize the evolution of the walker in the phase space. Following the Stratonovich-Weyl correspondence \cite{stratonovich_distributions_1957,varilly_moyal_1989,davis_wigner_2021}, the Wigner function for the SU(2) group can be defined as
	\begin{eqnarray}
		W(\theta, \phi) = \tr{\hat{\rho}_{\mathrm{w}} \hat{\Delta} (\theta, \phi)}{}, 
	\end{eqnarray}
	where the kernel can be written as
	\begin{eqnarray}
		\hat{\Delta}(\theta, \phi) &=& \sum_{m=-j}^j \Delta_{j,m} \ket{j, m; \mathbf{d}} \bra{j, m; \mathbf{d}}, \\
		\Delta_{j,m} &=& \sum_{l=0}^{2j} \frac{2l + 1}{2j + 1} \left< 
		\begin{array}{@{}cc|c@{}}
			j & l & j \\
			m & 0 & m
		\end{array}
		\right> .
	\end{eqnarray}
	Here $\ket{j, m; \mathbf{d}}$ is the Dicke basis along $\mathbf{d}=(\sin \theta \cos \phi, \sin \theta \sin \phi, \cos \theta)$ direction, which satisfies $\mathbf{d} \cdot \hat{\mathbf{J}} \ket{j, m; \mathbf{d}} = m \ket{j, m; \mathbf{d}}$. $\left<\begin{array}{@{}cc|c@{}}
		j & l & j \\
		m & 0 & m
	\end{array}\right>$ is the Clebsch-Gordan coefficient.

	The Wigner function satisfies the normalization relation
	\begin{eqnarray}
		\frac{2 J + 1}{4 \pi} \int_{0}^{\pi} \int_{-\pi}^{\pi} W(\theta, \phi) \sin \theta d\theta d\phi = 1,
	\end{eqnarray}
	and its marginal gives the probability distribution 
	\begin{eqnarray}
		P(\phi) = \frac{2 J + 1}{4 \pi} \int_{0}^{\pi}  W(\theta, \phi) \sin \theta d\theta .
	\end{eqnarray}
	Then one can define the standard deviation $\sigma$ (same as Eq. (\ref{eq:var})), with 
	\begin{eqnarray}
		\expect{\phi^l} = \int_{-\pi}^{\pi} P(\phi) \phi^l d\phi .
	\end{eqnarray}
	Here I focus on the short-time evolution when $\sigma$ is feasible to depict the quadratic enhancement of the quantum walk. For the long-time evolution, the Holevo standard deviation is more appropriate due to the periodic phase \cite{PhysRevA.78.042334}.
	
	Figure \ref{fig:PD} shows the quantum walk on the Bloch sphere for the first two steps. For comparison, the ideal quantum walk is also present. Initially, the walker+coin is described by $\ket{\psi(0)} = \ket{\phi_0}\otimes \ket{\uparrow}$. The walker can be regarded as a wave packet centered at $(\theta,\phi)=(\pi/2,0)$, as shown by the Wigner function in Fig. \ref{fig:PD} (a). The probability distributions for the spin coherent states (Fig. \ref{fig:PD} (d)) and ideal orthogonal states (Fig. \ref{fig:PD} (g))  are quite similar, except for the finite width in the former case. The finite width can be reduced by increasing the number o f spins $N$ in the spin cluster. After the first step $(k=1)$, one can easily prove that the walker's state becomes 
	\begin{eqnarray}
		\hat{\rho}_{\mathrm{w}} = \frac{1}{2} \left(\ket{\phi_1} \bra{\phi_1} + \ket{\phi_{-1}} \bra{\phi_{-1}}\right)  .
	\end{eqnarray}
	Initially localized wave packet propagates along opposite directions, which results in two uncorrelated wave packets, as shown in Figs. \ref{fig:PD} (b) and \ref{fig:PD} (e). After the second step, the walker's state takes the following form
	\begin{eqnarray}
		\hat{\rho}_{\mathrm{w}} &=& \frac{1}{4} \left(\ket{\phi_2} + \ket{\phi_0}\right) \left(\bra{\phi_2} + \bra{\phi_0}\right) \nonumber\\
		&+& \frac{1}{4} \left(\ket{\phi_0} - \ket{\phi_{-2}}\right) \left(\bra{\phi_0} - \bra{\phi_{-2}}\right) ,
	\end{eqnarray}
	which has two terms. Each term is composed of a superposition of two spin coherent states, that can be regarded as the spin cat state \cite{davis_wigner_2021,akhtar_sub-planck_2021}. The probability distributions for the spin coherent states (Fig. \ref{fig:PD} (f)) and ideal orthogonal states (Fig. \ref{fig:PD} (i))  are still quite similar. However, the Wigner function depicts more detailed structures, as shown in Fig. \ref{fig:PD} (c). There exist three wave packets, separated by fringes between them. The fringe is due to the interference between different coherent states, which is a distinguishing feature of the cat state. 
	
	\begin{figure}[htb]
		\centering
		\includegraphics[scale=0.95]{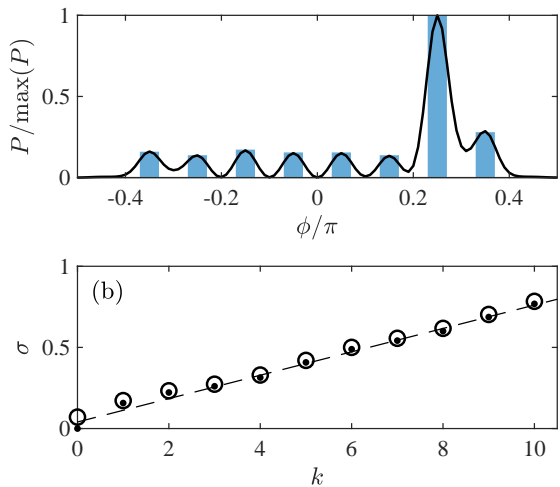} 
		\caption{Quantum walk on the Bloch sphere with $L=40$ and $N=200$. (a) Re-scaled probability distributions at $k=9$ for spin coherent states (black line) and ideal orthogonal states (blue bar); (b) standard deviations for spin coherent states (circle) and ideal orthogonal states (dot). A dashed line is plotted as a benchmark.}\label{fig:var}
	\end{figure}

	If more sites on the Bloch sphere are involved in the quantum walk, namely increasing $L$ and decreasing $\delta \phi$, a larger spin cluster with greater $N$ should be considered to make sure that the overlap ($\ref{eq:over_lap3}$) is small enough. Figure \ref{fig:var} exhibits the quantum walk on the Bloch sphere with $L=40$ sites for a spin cluster with $N=200$. As shown in Fig. \ref{fig:var} (a), the probability distributions for the spin coherent states and ideal orthogonal states are consistent  after $k=9$ steps, which exhibit more peaks. If the number of spins in the spin cluster decreases, the overlaps become larger, which can smear out the multi-peak structures \cite{Matjeschk_2012}. The standard deviation is depicted in Fig. \ref{fig:var} (b). Obviously, the standard deviation grows linearly with the number of steps  ($\sigma \propto k$), which is a characteristic feature of the quantum walk.
	
	\section{Conclusions and outlook} \label{sec:conclusion}
	
	The phase plane associating with the Heisenberg-Weyl group and the Bloch sphere associating with the SU(2) group are two well-known phase spaces. The quantum walk on the phase plane has been studied extensively based on the harmonic oscillator. However, little attention has been paid to the quantum walk on the Bloch sphere, to the best of my knowledge.
	
	In this paper, the discrete-time quantum walk in the phase space is generalized to the Bloch sphere. I focus on the spin cluster which serves as the walker, whereas other systems belonging to the SU(2) group follow the same pattern. The walker's locations on the Bloch sphere are determined by the spin coherent states, which are generally nonorthogonal. If the number of spins in the spin cluster increases, the overlap between different states decreases, which finally leads to the ideal quantum walk with orthogonal walker's states. To visualize the walking process on the Bloch sphere, the Wigner function is calculated. The probability distribution and the standard deviation are also calculated in virtue of the Wigner function, which confirm the quadratically growing variance, namely, $\sigma^2 \propto k^2$.
	
	The Bloch sphere serves as a new platform and offers more possibilities to study the quantum walk theoretically and experimentally. There are numerous related applications to be addressed. Here I just give three possibilities:  (1) Macroscopic superposition, such as the spin cat state, can be found during the walking process. The macroscopic superposed states may exhibit sub-Planck phase-space structures, which can be used to achieve the Heisenberg-limited sensitivity in weak-force measurements \cite{zurek_sub-planck_2001,PhysRevA.73.023803,akhtar_sub-planck_2021}. (2) One can extend the coin to include two spins. Two spins control the movement of the walker along a parallel and a meridian on the Bloch sphere separately, which leads to a two-dimensional quantum walk. (3) In the presence of the decoherence, the quantum walk tends to the classical random walk, which provides a new arena to study the quantum-to-classical transition \cite{travaglione_implementing_2002,PhysRevA.67.042305,brun_quantum_2003,karski_quantum_2009}.  The possible applications of quantum walk on the Bloch sphere and the influence of  decoherence deserve further studies, which are left to future research.

	%

\end{document}